\documentclass[
 reprint,
 amsmath,amssymb,
 aps,
]{revtex4-2}

\usepackage{graphicx}
\usepackage{dcolumn}
\usepackage{bm}
\usepackage{hyperref}
\hypersetup{
  breaklinks=true,   
  pdfusetitle=true,  
}

\begin{document}

\preprint{APS/123-QED}

\title{Improved Calculation of Acoustic Deformation Potentials from First Principles}%

\author{Patrick Williams}
\email{p.williams5@newcastle.ac.uk}
\author{Angela Dyson}%
\email{angela.dyson@newcastle.ac.uk}
\affiliation{School of Maths, Stats and Physics; Newcastle University, Newcastle upon Tyne, NE1 7RU, UK}%
\date{\today}

\begin{abstract}
  Using density functional theory (DFT) and density functional perturbation theory (DFPT), the band structure, phonon dispersion and electron phonon coupling matrix were calculated for silicon (Si), diamond and cubic boron nitride (cBN). From these, the acoustic deformation potential was calculated for multiple angles between the electron and phonon wave vectors and analytic expressions for the longitudinal and acoustic modes were fit to find an average deformation potential. The ability to calculate the deformation potential from first principles allows for the scattering rates to be determined without the use of lengthy empirical methods. For Si, the numerically calculated deformation potentials are in excellent agreement with what is seen in the literature. On the other hand, the deformation potentials calculated for diamond were found to be larger than what has been seen previously, however previous calculations of transport parameters in diamond report a large range of values for scattering parameters which may be due to assumptions made in each model. Excellent agreement was also seen between the value calculated for cBN and the literature, however there are no experimental results for cBN and so this value is compared against an estimate. This shows that scattering parameters can be calculated via first principles for materials with sparse experimental data, which in turn allows for increased confidence in the output of charge transport simulations of new and emerging materials.
\end{abstract}

\maketitle

\section{\label{sec:intro}Introduction}
It has been predicted that at least $70\%$ of electricity generated passes through power electronics \cite{Bose2013,Pearton2021,Kumar2022} and, with the advent of electric vehicles and the phasing out of fossil fuels, this is set to increase leading to increased pressure being placed on the electrical power sector. It is then necessary to develop more efficient devices to reduce losses and increase performance thus improving the energy efficiency of these devices. 

Using semiconductorsgmail with a  wider band gap results in a plethora of characteristics that make these materials ideal replacements for Si in power electronics. 4H-SiC and GaN have band gaps around 3.3 eV and 3.4 eV, respectively, roughly 3 times that of Si. Consequently, the breakdown voltage of these two materials is an order of magnitude greater than that of Si, resulting in devices that can be made $1/10$ the size while still sustaining the same applied voltage leading to lower switching and conduction losses, in turn allowing for higher switching speeds meaning smaller more efficient devices \cite{Pearton2021,Kumar2022,Takahashi2007_2}. Ultra-wide band gap semiconductors, as the name suggests, have even greater band gaps and so have the potential to be utilised in ways that wide-band gap materials cannot \cite{Tsao2018,Shur2022,Wong2021,Xu2022}. However, due to the nascent stage of development, experimental data on these materials may be sparse and so attention turns to computational methods to characterise these materials.

One important aspect of catagorising a material's properties is in modelling electron scattering methods induced by the interactions with phonons. This can be approximated using the deformation potential theory, developed by Bardeen and Shockley \cite{Bardeen1950} and generalised by Herring and Vogt \cite{Herring1956}. The deformation potential describes a shift in the band structure due to a perturbation in the lattice constant caused by phonon vibrations and is commonly employed to model non-polar phonon scattering mechanisms in Monte Carlo (MC) simulations to solve the Boltzmann Transport Equation. These scattering rates are calculated using Fermi's golden rule \cite{Schwinger2013} and generally take the form \cite{Ridley2013,Tomizawa1993},
\begin{equation}
  W(\mathbf{k}) = \kappa_{q}\Xi_{\nu}^{2}g(E_{\mathbf{k}'})
  \label{eq:scatter}
\end{equation}
where $\Xi_{\nu}$ is the deformation potential of phonon mode $\nu$, $\kappa_{\mathbf{q}}$ is a term that depends on the phonon mode and wavevector $\mathbf{q}$, and $g(E_{\mathbf{k}'})$ is the density of states (DOS) per unit volume at energy $E_{\mathbf{k}'}$.

For MC simulation, the deformation potentials are generally found empirically by fitting simulation results to experimental mobility or time of flight measurements \cite{Canali1975,Hammersberg2014,Majdi2016,Djurberg2022,Isberg2002,Nava1980,Tsukioka2001,Pernot2006}. As mentioned, research into UWBG semiconductors is still in the early stages, and so they are not as well understood as conventional semiconductors and lack the experimental results necessary for analytic or empirical modelling. The goal of this paper is then to improve upon methods for calculating the acoustic deformation potential via \textit{ab inito} methods. 

The paper is presented as follows: In section \ref{sec:adp}, the theory for how the acoustic deformation potential is calculated is laid out. This is followed by section \ref{sec:comp} where the specific computational methods employed are described. Section \ref{sec:Si_results} presents the results of the methods as applied to Si, chosen as it is a well characterised material and so the results can be compared to the plethora of studies previously conducted. Sections \ref{sec:diamond_results} and \ref{sec:cBN_restults} then detail the results of applying the methods to diamond and cubic boron nitride (cBN), respectively. These are both UWBG semiconductors that have been the subject of recent study, and were chosen due to their similarities in structure to Si. Section \ref{sec:conc} then draws everything together as a conclusion.

\section{\label{sec:theory}Theory and Methodology}
  \subsection{\label{sec:adp}Acoustic Deformation Potential}
    For acoustic phonon modes, the scattering rate takes the form
    \begin{equation}
      W_{ADP}(\mathbf{k}) = \frac{2\pi k_{B}T}{\hbar\rho v^{2}_{l}}\Xi_{ADP}^{2}g(E_{\mathbf{k}})
      \label{eq:ADP_scatter}
    \end{equation}
    where $\rho$ is the mass density and $v_{l}$ is the longitudinal velocity. This scattering rate arises from a perturbing potential that is proportional to the induced strain,
    \begin{equation}
      \mathcal{H}'_{ADP} = \Xi_{ADP} \nabla \cdot \mathbf{u}_{\mathbf{q}}(\mathbf{R}, t)
      \label{eq:pert_pot}
    \end{equation}
    where $\mathcal{H}'_{adp}$ is the perturbing potential; $\mathbf{u}_{\mathbf{q}}(\mathbf{R}, t)$ is the displacement of the nuclei due to phonon with wavevector $\mathbf{q}$; and the proportionality constant, $\Xi_{adp}$, is the acoustic deformation potential. The displacement of the nuclei from the lattice position $\mathbf{R}$ can be expanded as a Bloch wave with amplitude $A_{\mathbf{q}}$, 
    \begin{equation}
      \mathbf{u}_{\mathbf{q}}(\mathbf{R}, t) = A_{\mathbf{q}}e^{i(\mathbf{q}\cdot\mathbf{R}-\omega t)}+A_{\mathbf{q}}e^{-i(\mathbf{q}\cdot\mathbf{R}-\omega t)}
      \label{eq:displacement}
    \end{equation}
    where, $\omega$ is the frequency of phonon with wavevector $\mathbf{q}$. For a small displacement, this strain can be approximated as the displacement multiplied by the magnitude of the phonon wavevector,
    \begin{equation}
      \nabla \cdot \mathbf{u}_{\mathbf{q}}(\mathbf{R}, t) \approx |\mathbf{q}|\mathbf{u}_{\mathbf{q}}(\mathbf{R}, t)
      \label{eq:strain_approx}
    \end{equation}

    The perturbing potential can then be found by considering its relation to the electron-phonon coupling matrix as
    \begin{equation}
      \sqrt{\frac{2m_{0}\omega}{\hbar}}g_{m,n}^{\nu}(\mathbf{k},\mathbf{q})=\langle\psi_{m,\mathbf{k}+\mathbf{q}}|\delta_{\mathbf{u}}\mathcal{H}'_{adp}|\psi_{n,\mathbf{k}}\rangle
      \label{eq:el_ph}
    \end{equation}
    Here, $m_0$ is the mass of the unit cell; $|\psi_{n,\mathbf{k}}\rangle$ and $|\psi_{m,\mathbf{k}+\mathbf{q}}\rangle$ are the initial and final electronic wavefunctions in band $n$ with state $\mathbf{k}$ and in band $m$ with state $\mathbf{k}+\mathbf{q}$, respectively; 
    and $\delta_{\mathbf{u}}\mathcal{H}'_{adp}$ is the linear change in the perturbing potential due to the displacement $\mathbf{u}_{\mathbf{q}}(\mathbf{R}, t)$. The term $g_{m,n}^{\nu}(\mathbf{k},\mathbf{q})$ is the element of the electron-phonon coupling matrix that describes the transition of an electron from state $\mathbf{k}$ in band $n$ to state $\mathbf{k}+\mathbf{q}$ in band m due to a phonon in mode $\nu$ with momentum $\mathbf{q}$. For small $\mathbf{q}$, the overlap $\langle\psi_{m,\mathbf{k}+\mathbf{q}}|\psi_{n,\mathbf{k}}\rangle$ can be approximated as unity and therefore the perturbing potential can be approximated as,
    \begin{equation}
      \mathcal{H}'_{adp} = \sqrt{\frac{2m_{0}\omega}{\hbar}}g_{m,n}^{\nu}(\mathbf{k},\mathbf{q})\cdot \mathbf{u}_{\mathbf{q}}(\mathbf{R}, t)
      \label{eq:pert_pot_approx}
    \end{equation}
    Substituting equations \ref{eq:strain_approx} and \ref{eq:pert_pot_approx} into equation \ref{eq:pert_pot} and rearranging means the acoustic deformation potential can be found as,
    \begin{align}
      \Xi_{adp} &= \sqrt{\frac{2m_{0}\omega}{\hbar}}\frac{g_{m,n}^{\nu}(\mathbf{k},\mathbf{q})\cdot \mathbf{u}_{\mathbf{q}}(\mathbf{R}, t)}{|\mathbf{q}|\mathbf{u}_{\mathbf{q}}(\mathbf{R}, t)} \nonumber\\ &= \sqrt{\frac{2m_{0}\omega}{\hbar}}\frac{g_{m,n}^{\nu}(\mathbf{k},\mathbf{q})}{|\mathbf{q}|}
      \label{eq:adp}
    \end{align}
    In the limit as $|\mathbf{q}|\rightarrow0$, the acoustic deformation potential is then defined as the gradient of the electron-phonon coupling matrix with respect to $|\mathbf{q}|$.
    \begin{equation}
      \Xi_{adp} = \frac{d}{d|\mathbf{q}|}\left|\sqrt{\frac{2m_{0}\omega}{\hbar}}g_{m,n}^{\nu}(\mathbf{k},\mathbf{q})\right|
      \label{eq:adp_2}
    \end{equation}

  \subsection{\label{sec:comp}Computational Methods}
  
    To 
    calculate the acoustic deformation potential, DFT and DFPT calculations were done using the software package Quantum Espresso \cite{Giannozzi2009,Giannozzi2017,Giannozzi2020} along with the extension EPW \cite{Ponce2016,Lee2023} to calculate the electron-phonon coupling matrix via the Wannier-Fourier interpolation method.
    The nuclei were modelled using norm-conserving pseudopotentials \cite{Hartwigsen1998, Hamann2013, Schlipf2015} with the PBE-GGA exchange-correlation functional. The Kohn-Sham eigenfunctions are represented using a planewave basis set with a cut-off energy of $100.0$ Ry. The Brillouin zone was sampled with a $20\times20\times20$ Monkhorst-Pack grid \cite{Monkhorst1976} for the $k$-points during the initial Quantum Espresso structure relaxation and self-consistent field calculations. This was then used in conjuncture with a $10\times10\times10$ Monkhorst-Pack grid for the $q$-points during the phonon dispersion calculations and EPW calculations of the electron-phonon coupling matrix.

    The longitudinal and transverse acoustic phonon deformation potential can then be calculated for the conduction band minimum using equation \ref{eq:adp} for a given phonon wavevector, however, these deformation potentials have a dependence on the angle between $\mathbf{k}$ and $\mathbf{q}$ \cite{Ridley2013,Herring1956} given by,
    \begin{subequations}
      \begin{align}
          \Xi_{LA}(\theta) &= \Xi_{d}+\Xi_{u} cos^{2}(\theta) \label{eq:LA_angle}\\
          \Xi_{TA}(\theta) &= \Xi_{u} sin(\theta)cos(\theta)  \label{eq:TA_angle}
      \end{align}
  \end{subequations}
  where $\Xi_{LA}(\theta)$ and $\Xi_{TA}(\theta)$ are the angle dependent deformation potentials for the longitudinal acoustic and the combined transverse acoustic phonon modes, respectively; $\Xi_{d}$ and $\Xi_{u}$ are the dilatation and uniaxial deformation potential constants, respectively; and $\theta$ is the polar angle between $\mathbf{k}$ and $\mathbf{q}$. Equation \ref{eq:adp_2} can then be used to calculate the longitudinal and transverse acoustic deformation potentials for a range of polar angles between $0$ and $\pi$, and then by fitting equations \ref{eq:LA_angle} and \ref{eq:TA_angle}, values can be found for $\Xi_{d}$ and $\Xi_{u}$ can be determined. A value for the longitudinal and transverse deformation potentials, independent of the angle between $\mathbf{k}$ and $\mathbf{q}$, can be found as the average of the integral of the square of equations \ref{eq:LA_angle} and \ref{eq:TA_angle} between $0$ and $\pi$. These averages are given by,
  \begin{subequations}
      \begin{align}
          \Xi_{LA}^{2} = \Xi_{d}^{2}&+\Xi_{d}\Xi_{u}+\frac{3}{8}\Xi_{u}^{2}\label{eq:LA_average}\\
          \Xi_{TA}^{2} &= \frac{\Xi_{u}^{2}}{8} \label{eq:TA_average}
      \end{align}
  \end{subequations}
  Generally, only a single value is given for the acoustic deformation potential and so the contribution of the longitudinal and transverse acoustic deformation potentials can then be combined as,
  \begin{equation}
      \Xi_{ADP}^{2} = \Xi_{LA}^{2} + \left(\frac{v_{l}^{2}}{v_{t}^{2}}\right)\Xi_{TA}^{2}
      \label{eq:ADP_contribution}
  \end{equation}
  where $v_{l}$ and $v_{t}$ are the longitudinal and transverse velocities of sound, respectively.

  As well as varying the polar angle, the azimuthal angle can be varied to see how this impacts the deformation potential. It is then important to consider how the deformation potentials are combined in the case of finding the average and also in finding $\Xi_{TA}(\theta)$ for equation \ref{eq:TA_angle} as this is the combination of both transverse modes. By considering equation \ref{eq:ADP_scatter}, it can be seen that the addition of two acoustic scattering rates has the impact of summing the square of the deformation potential, 
  \begin{equation}
    \Xi_{tot} = \sqrt{\sum_{i=1}^{n}\Xi_{i}^{2}}
    \label{eq:DP_combine}
  \end{equation}
  and the average deformation potential is found as the root mean square of the contributions to the deformation potential,
  \begin{equation}
    \Xi_{tot} = \sqrt{\frac{\sum_{i=1}^{n}\Xi_{i}^{2}}{n}}
    \label{eq:DP_ems}
  \end{equation}
  Following these methods, the acoustic deformation potential is calculated via first principles for Si, diamond and cBN in the following sections.

\section{\label{sec:Results}Results}
  The data used to generate the following figures are available at reference \cite{Williams2025}.

  \subsection{\label{sec:Si_results}Silicon}
  
    DFT calculations were performed on Si using the parameters and pseudopotentials detailed above. The relaxed structure was found to have a lattice constant of $5.461$ \AA, a percentage difference of $0.57\%$ from the experimental value \cite{Hom1975}. The bulk modulus and the cohesive energy per atom were also found to have a percentage error of $10.6\%$ \cite{Hopcroft2010} and $0.065\%$ \cite{Kittel2005}, respectively, when compared to experimental results.

    \begin{figure}[h]
      \centering
      \includegraphics[width = 0.62\linewidth,angle=-90]{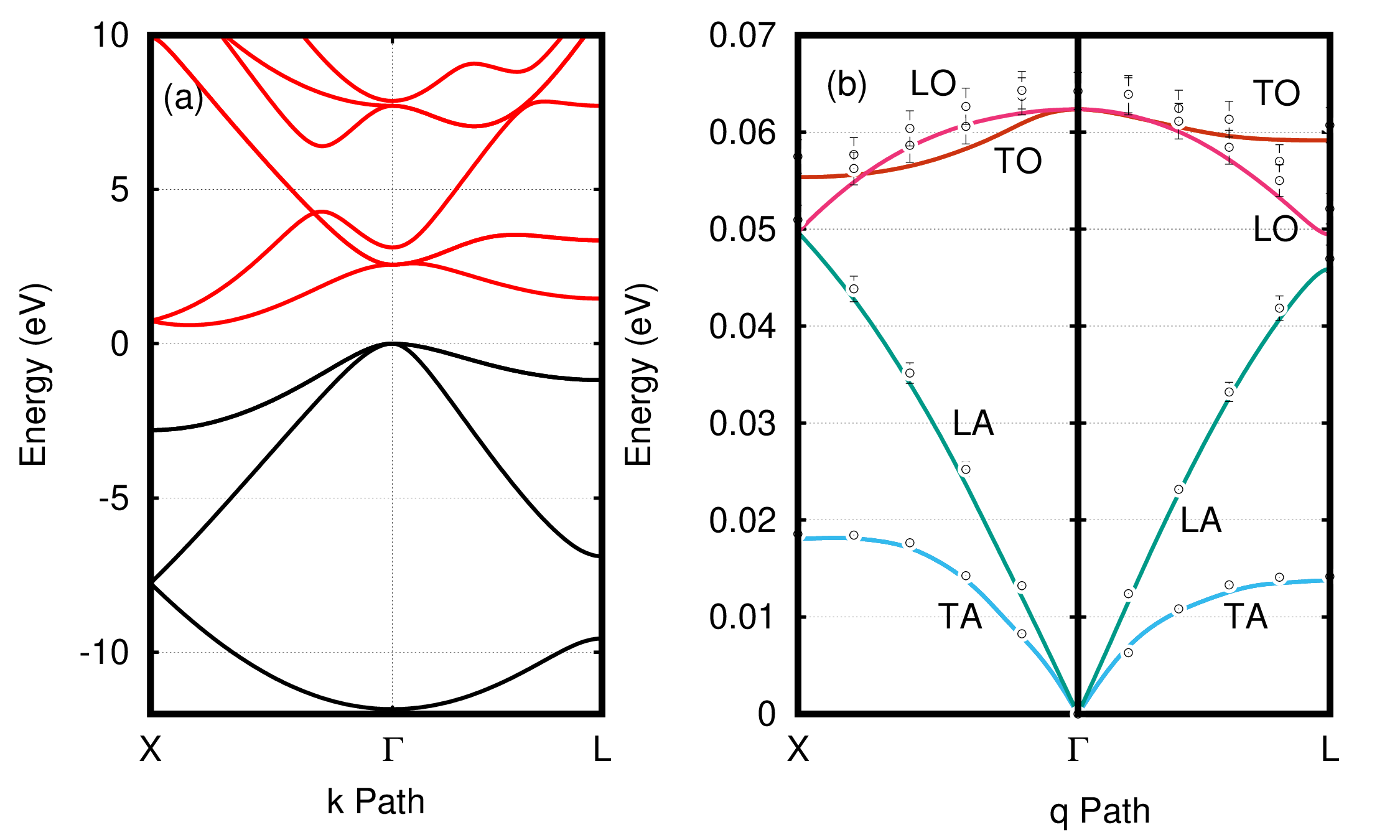}
      \caption{(a) The electronic band structure of silicon as calculated using DFT. The conduction band minimum is located at $\sim 84\%$ along the $\Delta$ direction. (b) The phonon dispersion of silicon calculated using DFT (solid lines) and experimentally (points). Experimental data taken from reference \cite{Dolling_1962}.}
      \label{fig:Si_bands}
    \end{figure}
    
    Figure \ref{fig:Si_bands} then shows the electronic band structure and the phonon dispersion calculated for Si once the relaxed structure was found. The band structure shows the conduction band minimum being located at roughly $84\%$ of the distance between the $\Gamma$ and $X$ high symmetry points, in agreement with what is seen in the literature. However, the band gap is calculated to be $0.604$eV which is just greater than half the experimentally determined value \cite{Tang1983}. This is due to the well known band gap problem in DFT and should not impact the results presented here. The phonon dispersion shows good agreement between the numerically calculated phonon energies and those collected experimentally \cite{Dolling_1962}, of particular note and importance here is the agreement between the experimentally and numerically calculated acoustic phonon modes.

    The longitudinal and transverse acoustic deformation potentials were found using equation \ref{eq:adp_2} with the polar angle between the phonon vector and the longitudinal axis of the $k_{z}$ conduction band minimum valley found at twenty evenly spaced angles from $0$ to $\pi$. For each polar angle, the azimuthal angle was varied and the average of the deformation potentials found from eight evenly spaced points between $0$ and $2\pi$. 

    \begin{figure}[h]
        \centering
        \includegraphics[width = 0.62\linewidth,angle=-90]{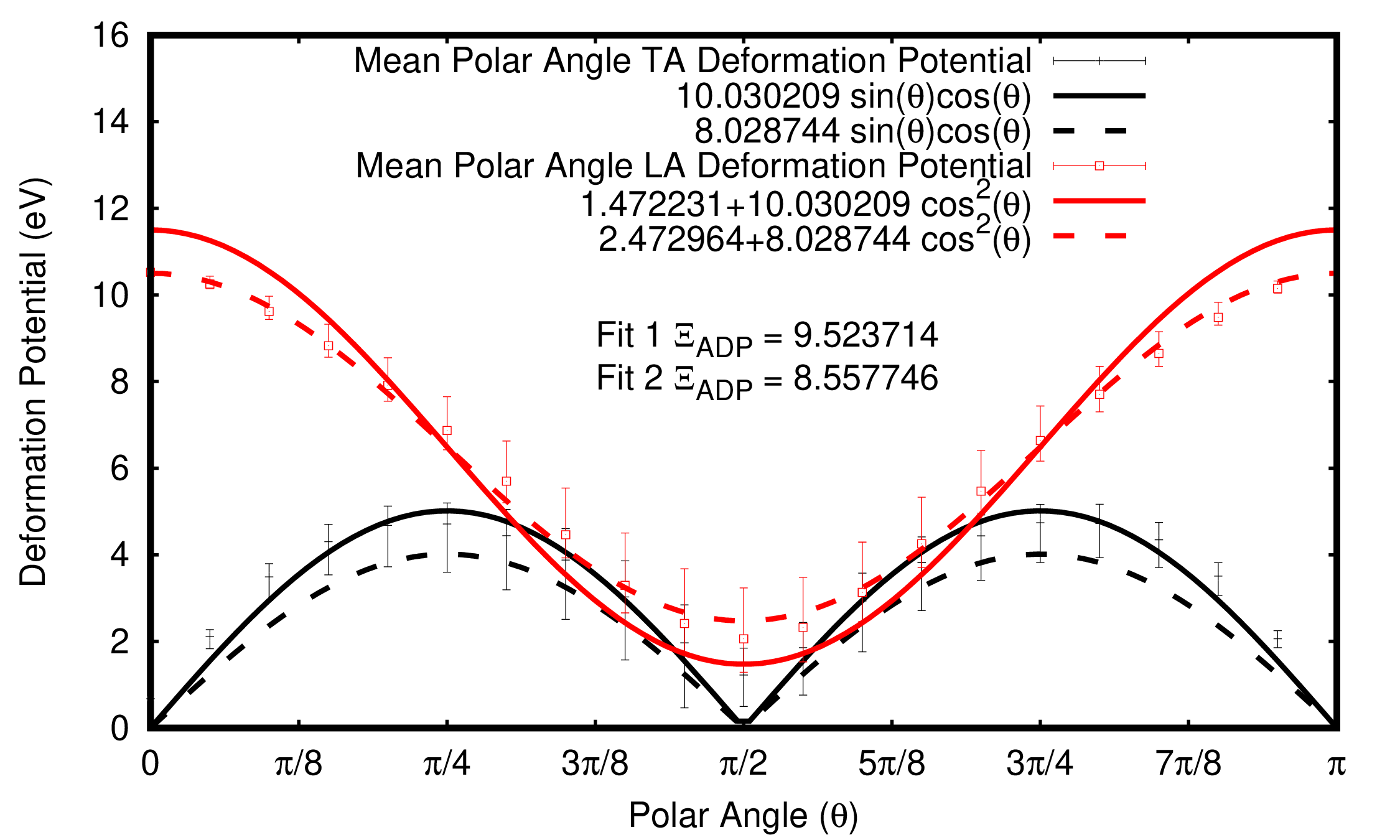}
        \caption[Polar Dependence of the Acoustic Deformation Potential in Si.]{Polar dependence of the longitudinal (red) and transverse (black) acoustic deformation potential in the first conduction band valley of Si calculated via quantum espresso (points) and equations \ref{eq:LA_angle} and \ref{eq:TA_angle}. The bars indicate the range of values found by varying the azimuthal angle for a given polar angle.}
        \label{fig:Si_HV}
    \end{figure}
    \begin{table*}
      \caption{Comparison of the acoustic deformation potential and effective mass for Si quoted in literature and values presented in this work.}
      \label{tbl:Si_ADP}
      \begin{ruledtabular}
        \begin{tabular}{ccccccc}
           & $\Xi_{d} (eV)$ & $\Xi_{u} (eV)$ & $\Xi_{ADP} (eV)$ & $m_{l}/m_{0}$ & $m_{t}/m_{0}$ & $m_{DOS}/m_{0}$\\
        \colrule
          Canali et al. \cite{Canali1975} & - & - & 9.0 & 0.9163 & 0.1905 & 0.322\\
          Tomizawa \cite{Tomizawa1993} & - & - & 9.0 & 0.92 & 0.19 & 0.32\\
          Fischetti and Laux \cite{Fischetti1996} & 1.1 & 10.5 & 9.7 & 0.9163 & 0.1905 & 0.322\\          
          Li et al. \cite{Li2021}& 1.01 & 8.84 & 8.2 & -& -&-\\
          Yang et al. \cite{Yang2024}& 0.8 & 9.0 & 8.2 & -& -&-\\
          Fit 1 & 1.47 & 10.03 & 9.52 & 0.958 & 0.204 & 0.341\\
          Fit 2 & 2.47 & 8.03 & 8.55 & 0.958 & 0.204 & 0.341
        \end{tabular}
        \end{ruledtabular}
    \end{table*}
    Figure \ref{fig:Si_HV} shows the angular dependence of the acoustic deformation potentials, with the red plots showing the angular dependence of the longitudinal acoustic deformation potential and the black plots showing the sum of the deformation potential of the two transverse acoustic modes. The points show the average deformation potential for a given polar angle and the bars show the range of values calculated for the deformation potential with a given polar angle and varied azimuthal angle. The lines in figure \ref{fig:Si_HV} show the fit of equations \ref{eq:LA_angle} and \ref{eq:TA_angle} to the calculated deformation potentials. The solid lines show the case where $\Xi_{u}$ and $\Xi_{d}$ were found separately by first fitting equation \ref{eq:TA_angle} to the average polar angle transverse acoustic deformation potential to find the value of $\Xi_{u}$ and then fixing this while fitting equation \ref{eq:LA_angle} to the average polar angle longitudinal acoustic deformation potential to find the value of $\Xi_{d}$. In this method of fitting the parameters separately, the uniaxial and dilatation deformation potentials were found to be $\Xi_{u}=10.03$eV and $\Xi_{d}=1.47$eV.
    The dashed lines show the case where $\Xi_{u}$ and $\Xi_{d}$ were found together by fitting equation \ref{eq:LA_angle} to the average polar angle longitudinal acoustic deformation potential. In this case, the uniaxial and dilatation deformation potentials were found to be $\Xi_{u}=8.03$eV and $\Xi_{d}=2.47$eV. The black dashed line is then the plot of equation \ref{eq:TA_angle} where $\Xi_{u}=8.03$eV.
    
    Values for the longitudinal and transverse acoustic deformation potential can be found using equations \ref{eq:LA_average} and \ref{eq:TA_average}. The contributions of the longitudinal acoustic and transverse acoustic modes can be combined into a single value using equation \ref{eq:ADP_contribution} with $v_{l} = 9.04\times 10^{3}$ m/s and $v_{t} = 5.34\times 10^{3}$ m/s \cite{Urban_PhD_Thesis,Li2021,Jacoboni1977}. The acoustic deformation potential found by fitting equation \ref{eq:TA_angle} to the mean polar angle transverse acoustic deformation potential to find $\Xi_{u}$ and then fitting \ref{eq:LA_angle} to the mean polar angle longitudinal acoustic deformation potential to find $\Xi_{d}$, referred to as fit 1, was $\Xi_{ADP}=9.52$eV. The acoustic deformation potential found by fitting equation \ref{eq:LA_angle} to the mean polar angle longitudinal acoustic deformation potential to find both $\Xi_{d}$ and $\Xi_{u}$, referred to as fit 2, was $\Xi_{ADP}=8.56$eV. 

    Table \ref{tbl:Si_ADP} shows a comparison of the deformation potentials and effective masses for the first conduction band of Si as quoted in the literature and calculated here numerically. The combined deformation potential ($\Xi_{ADP}$) is found using equations \ref{eq:LA_average}, \ref{eq:TA_average} and \ref{eq:ADP_contribution}; and the DOS effective mass is found as the geometric mean of the longitudinal and transverse effective masses ($m^{*}_{DOS}=\sqrt[3]{m_{l}^{*}m_{t}^{*}m_{t}^{*}}$), where they are given.

    As can be seen from the fourth column in table \ref{tbl:Si_ADP}, the values calculated for the acoustic deformation potential give good agreement with what is seen in the literature. Li et al. \cite{Li2021} and Yang et al. \cite{Yang2024} also calculate the acoustic deformation potential from the electron phonon coupling matrix and so good agreement is expected, however, these references calculate the uniaxial and dilatation deformation potentials at the points where one or the other goes to 0 in equations \ref{eq:LA_angle} and \ref{eq:TA_angle} whereas the method employed here uses the entire range of polar angles.

    The fifth and sixth columns of table \ref{tbl:Si_ADP} show the longitudinal and transverse effective masses at the conduction band minimum as multiples of the electron rest mass. The final column shows the DOS effective mass which is generally used to generate the DOS employed in MC simulations. The effective masses presented here were calculated by fitting a second order polynomial equation to the band structure close to the conduction band minimum parallel ($m_{l}$) and perpendicular ($m_{t}$) to the $\Delta$ direction. As can be seen from table \ref{tbl:Si_ADP}, the numerically calculated effective masses compare well with those given in the literature, calculated experimentally through cyclotron resonance \cite{Hensel1965}; with a percentage error of $4.6\%$ in $m_{l}$ and $7.1\%$ in $m_{t}$. By considering equation \ref{eq:ADP_scatter}, it can be seen that this overestimate in the effective mass would lead to an increase in the acoustic scattering at lower energies where the acoustic scattering dominates. Despite this overestimate, the methods employed here give a good estimate for the scattering parameters for Si with minimal reliance on experimental results.

  \subsection{\label{sec:diamond_results}Diamond}

    DFT calculations were performed on diamond using the parameters and pseudopotentials detailed above. The relaxed structure was found to have a lattice constant of $3.57$ \AA, a percentage difference of $0.13\%$ from the experimental value \cite{Ohring2002, Pan1994}. The bulk modulus and the cohesive energy per atom were also found to have a percentage error of $2.4\%$ \cite{McSkimin1972} and $2.2\%$ \cite{Kittel2005}, respectively, when compared to experimental results.

    \begin{figure}[h]
      \centering
      \includegraphics[width = 0.62\linewidth,angle=-90]{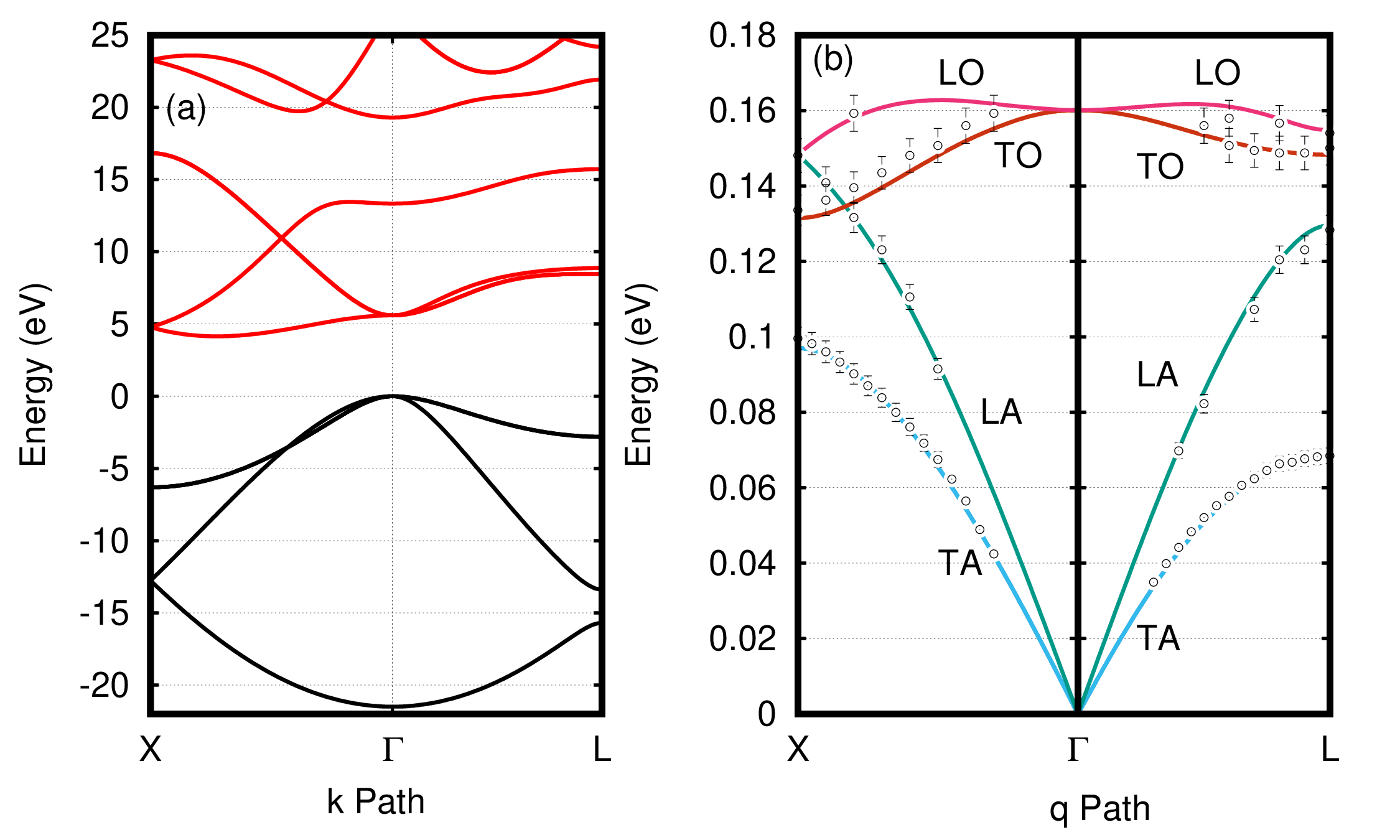}
      \caption{(a) The electronic band structure of diamond as calculated using DFT. The conduction band minimum is located at $~82\%$ along the $\Delta$ direction. (b) The phonon dispersion of diamond calculated using DFT (solid lines) and experimentally (points). Experimental data taken from reference \cite{Warren1965}.}
      \label{fig:diamond_bands}
    \end{figure}

    Figure \ref{fig:diamond_bands} then shows the electronic band structure and the phonon dispersion calculated for diamond using the relaxed structure. The band structure shows the conduction band minimum being located at roughly $73\%$ of the distance between the $\Gamma$ and $X$ high symmetry points, in agreement with what is seen in the literature. Once again the band gap is found to be underestimated at $4.15$eV, $24\%$ smaller than the experimentally measured value of $5.47$ eV \cite{Yang2021,Saslow1968}.
    The phonon dispersion shows good agreement between the numerically calculated phonon energies and those collected experimentally \cite{Warren1965}, of particular note is the agreement between the experimentally and numerically calculated acoustic phonon modes.
    
    As with Si, the longitudinal and transverse acoustic deformation potentials were found using equation \ref{eq:adp_2} with the polar angle between the phonon vector and the longitudinal axis of the $k_{z}$ conduction band minimum valley found at twenty evenly spaced angles from $0$ to $\pi$. For each polar angle, the azimuthal angle was varied and the average of the deformation potential found from eight evenly spaced points between 0 and $2\pi$.

    \begin{figure}[ht]
        \centering
        \includegraphics[width = 0.62\linewidth,angle=-90]{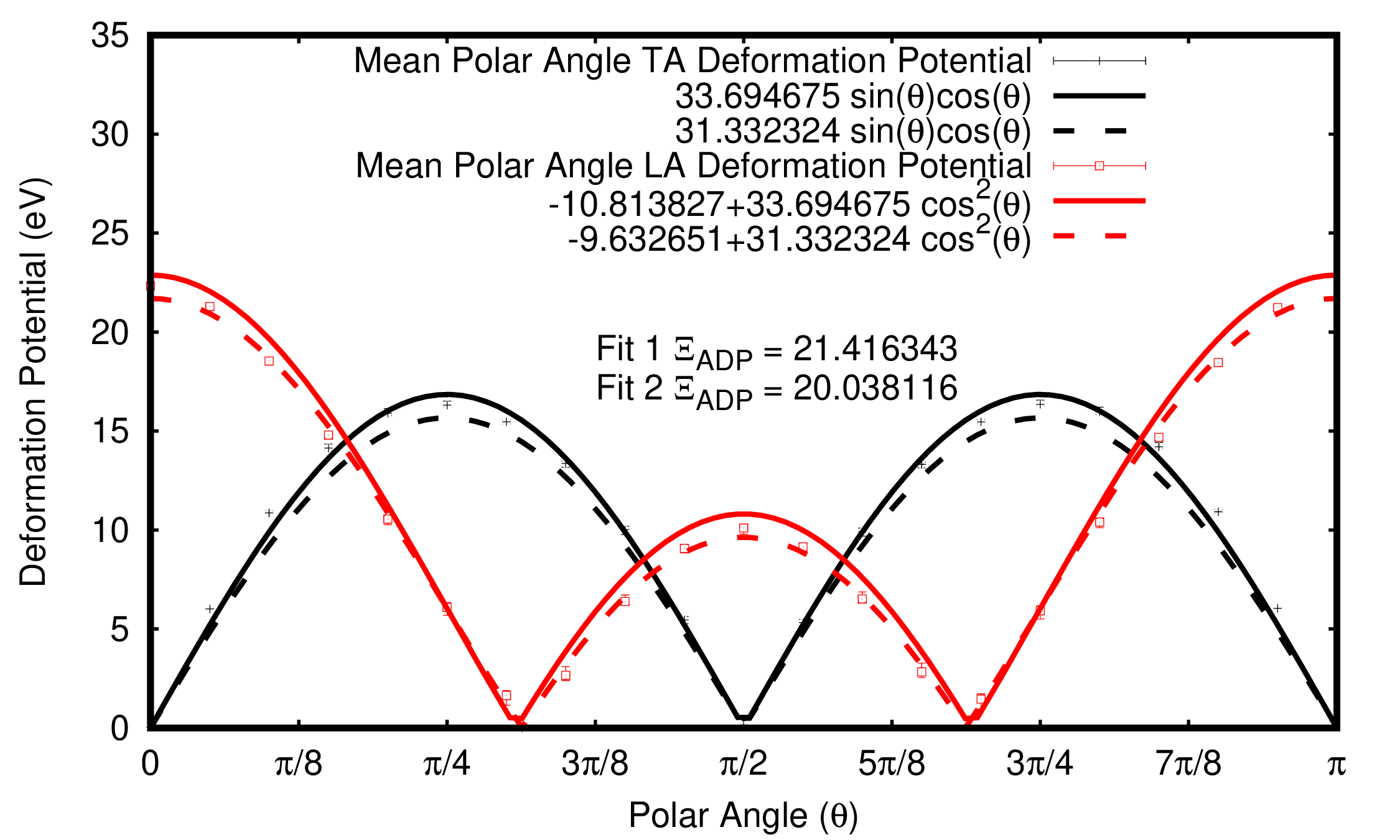}
        \caption[Polar Dependence of the Acoustic Deformation Potential in diamond.]{Polar dependence of the longitudinal (red) and transverse (black) acoustic deformation potential in the first conduction band valley of diamond calculated via quantum espresso (points) and equations \ref{eq:LA_angle} and \ref{eq:TA_angle}. The bars indicate the range of values found by varying the azimuthal angle for a given polar angle.}
        \label{fig:diamond_HV}
    \end{figure}
    Figure \ref{fig:diamond_HV} shows the angular dependence of the acoustic deformation potentials for the longitudinal, red, and the sum of the two transverse, black, acoustic phonon modes. The points show the average deformation potential for a given polar angle and the bars show the range of values calculated while varying the azimuthal angle for a given polar angle. The lines in figure \ref{fig:diamond_HV} show the fit of equations \ref{eq:LA_angle} and \ref{eq:TA_angle} to the calculated deformation potentials. Unlike Si, there is very little variation in the acoustic deformation potential calculated at different azimuthal angles, resulting in very narrow bars. 

    The lines in figure \ref{fig:diamond_HV} show the fit of equations \ref{eq:LA_angle} and \ref{eq:TA_angle} to the calculated deformation potentials. The solid lines show the case where $\Xi_{u}$ and $\Xi_{d}$ were found separately by first fitting equation \ref{eq:TA_angle} to the average polar angle transverse acoustic deformation potential to find the value of $\Xi_{u}$ and then fixing this while fitting equation \ref{eq:LA_angle} to the average polar angle longitudinal acoustic deformation potential to find the value of $\Xi_{d}$. The uniaxial and dilatation deformation potentials were found to be $\Xi_{u}=33.69$eV and $\Xi_{d}=-10.81$eV.
    The dashed lines show the case where $\Xi_{u}$ and $\Xi_{d}$ were found together by fitting equation \ref{eq:LA_angle} to the average polar angle longitudinal acoustic deformation potential. The uniaxial and dilatation deformation potentials were found to be $\Xi_{u}=31.33$eV and $\Xi_{d}=-9.63$eV. The black dashed line is then the plot of equation \ref{eq:TA_angle} for $\Xi_{u}=31.33$eV.

    Values for the longitudinal and transverse acoustic deformation potential can then be found using equations \ref{eq:LA_average} and \ref{eq:TA_average}, these contributions of the longitudinal and transverse acoustic modes can be combined into a single value using equation \ref{eq:ADP_contribution} with $v_{l} = 18.038\times 10^{3}$ m/s and $v_{t} = 12.834\times 10^{3}$ m/s \cite{Wang2004}. The acoustic deformation potential found by fitting equation \ref{eq:TA_angle} to the mean polar angle transverse acoustic deformation potential to find $\Xi_{u}$ and then fitting \ref{eq:LA_angle} to the mean polar angle longitudinal acoustic deformation potential to find $\Xi_{d}$, referred to as fit 1, was $\Xi_{ADP}=21.42$eV. The acoustic deformation potential found by fitting equation \ref{eq:LA_angle} to the mean polar angle longitudinal acoustic deformation potential to find both $\Xi_{d}$ and $\Xi_{u}$, referred to as fit 2, was $\Xi_{ADP}=20.04$eV. 

    \begin{table*}
      \caption{Comparison of the acoustic deformation potentials and effective masses for diamond quoted in literature and values presented in this work.}
      \label{tbl:diamond_ADP}
      \begin{ruledtabular}
        \begin{tabular}{ccccccc}
           & $\Xi_{d} (eV)$ & $\Xi_{u} (eV)$ & $\Xi_{ADP} (eV)$ & $m_{l}/m_{0}$ & $m_{t}/m_{0}$ & $m_{DOS}/m_{0}$\\
        \colrule
          Nava et al. \cite{Nava1980} & - & - & 8.7 & 1.4 & 0.36 & 0.57\footnote{Empirically calculated by fitting simulation to experimental results.}\\
          Tsukioka \cite{Tsukioka2001,Tsukioka2009} & - & - & 8.8,8.0 & - & - & 0.65\footnote{Calculated by fitting to DFT DOS}\\
          Pernot et al. \cite{Pernot2006} & - & - & 17.7 & 1.81 & 0.306 & 0.55\footnote{Taken from reference \cite{GHEERAERT2001}.} \\
          Isberg et al. \cite{Isberg2021}& - & - & 15 & 1.15 & 0.22 & 0.38\footnote{Taken from reference \cite{Lofas2011}.}\\
          Hammersberg et al. \cite{Hammersberg2014} & - & - & 12.0 & 1.56 & 0.28 & 0.50\footnotemark[5] \footnotetext[5]{Taken from reference \cite{Naka2013}.}\\  
          Majdi et al. \cite{Majdi2016} & - & - & 11.5 & 1.56 & 0.28 & 0.50\footnotemark[5]\\
          Djurberg et al. \cite{Djurberg2022} & -5.7 & 18.5 & 11.8 & 1.56 & 0.28 & 0.50\footnotemark[5]\\
          Fit 1 & -10.81 & 33.69 & 21.42 & 1.64 & 0.29 & 0.52\\
          Fit 2 & -9.63 & 31.33 & 20.04 & 1.64 & 0.29 & 0.52
        \end{tabular}
        \end{ruledtabular}
    \end{table*}
    Table \ref{tbl:diamond_ADP} shows a comparison of the deformation potentials and effective masses for the first conduction band of diamond as quoted in the literature and calculated here numerically. In the case of Djurberg et al. \cite{Djurberg2022} where only the uniaxial and dilatation deformation potentials were given, the combined acoustic deformation potential was calculated using equations \ref{eq:LA_average}, \ref{eq:TA_average} and \ref{eq:ADP_contribution}. The DOS effective mass is found as the geometric mean of the longitudinal and transverse effective masses ($m^{*}_{DOS}=\sqrt[3]{m_{l}^{*}m_{t}^{*}m_{t}^{*}}$) where they are given. 
    
    As can be seen, there are a large range of values given for the acoustic deformation potential, with the lower end giving similar values to Si, table \ref{tbl:Si_ADP}; and the upper end giving values more than double that. The results of both fitting methods give a combined acoustic deformation potential greater than what is seen in the literature with fit 2 being slightly closer. It is also interesting to note that the dilatation and uniaxial deformation potentials from both fits are almost double what is given by Djurberg et al. \cite{Djurberg2022}. This large range in values is also seen in the effective mass shown in table \ref{tbl:diamond_ADP} with the values presented depending on the method of calculation, which are shown in the table footnote.

    This discrepancy may then be due to the choice of parameters employed in the calculation. As seen in equation \ref{eq:ADP_scatter}, the acoustic phonon scattering rate depends on the acoustic deformation potential and the DOS which in turn depends on the effective mass. This means that these can act as free parameters when fitting the model to the experimental results and so the value chosen for the effective mass has an impact on the empirically determined deformation potential, chosen to obtain the best possible agreement with experimental results. Although the acoustic deformation potential is larger than values previously reported, the effective mass is in good agreement with the values used by references \cite{Hammersberg2014,Majdi2016,Djurberg2022}, which was found experimentally via cyclotron resonance \cite{Naka2013}. The percentage error between the experimentally calculated effective mass and what was calculated here was then $5.1\%$ in $m_{l}$ and $3.6\%$ in $m_{t}$.
    
    As the method presented here is an \textit{ab initio} method, it requires minimal experimental results. This led to effective masses in good agreement with experimental results \cite{Naka2013}; and acoustic deformation potentials found without comparison to time of flight measurements as used in references \cite{Hammersberg2014,Majdi2016,Djurberg2022,Isberg2002}, fitting to the electron mobility as in references \cite{Nava1980,Tsukioka2001,Pernot2006}.

  \subsection{\label{sec:cBN_restults}Cubic Boron Nitride}
    DFT calculations were performed on cBN using the parameters and pseudopotentials detailed above. The relaxed structure was found to have a lattice constant of $3.59$ \AA, a percentage difference of $0.81\%$ from the experimental value \cite{Levinshtein2001, Vel1991}. The bulk modulus and the cohesive energy per atom were also found to have a percentage error of $3.9\%$ \cite{Levinshtein2001} and $2.9\%$ \cite{lam1990}, respectively, when compared to experimental results. 

    \begin{figure}[h]
      \centering
      \includegraphics[width = 0.62\linewidth,angle=-90]{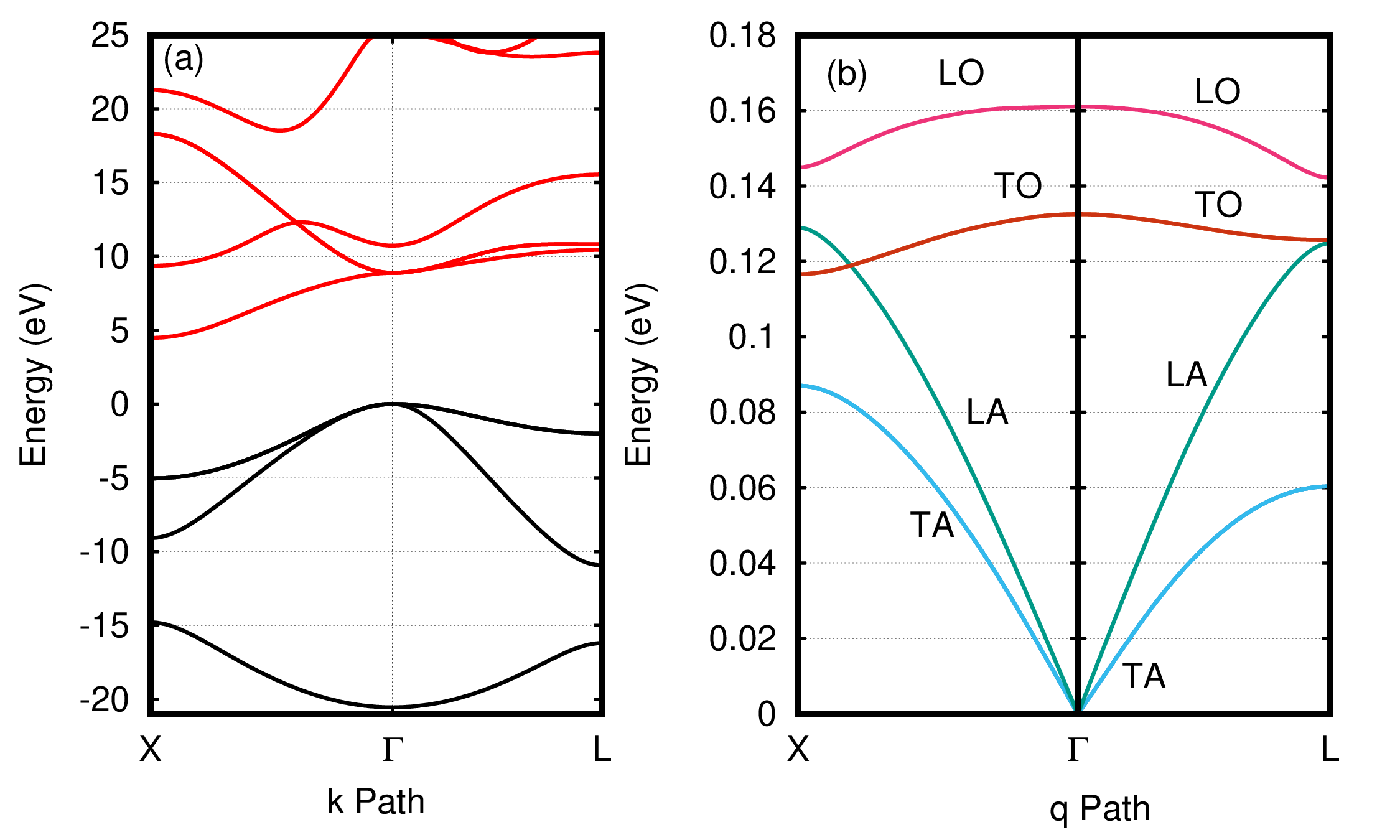}
      \caption{(a) The electronic band structure of cBN as calculated using DFT. The conduction band minimum is located at $~82\%$ along the $\Delta$ direction. (b) The phonon dispersion of cBN calculated using DFT.}
      \label{fig:cBN_bands}
    \end{figure}

    Figure \ref{fig:cBN_bands} shows the electronic band structure and the phonon dispersion calculated for cBN using the relaxed structure. The band structure shows the conduction band minimum being located at the $X$ high symmetry points, in agreement with what is seen in the literature. The band gap is found to be underestimated at $4.49$eV, $25\%$ smaller than the experimentally measured value of $6.1$ eV \cite{Levinshtein2001, Miyata1989}. The phonon dispersion shows good agreement with what is seen in other numerical calculations for cBN \cite{Zhu2023, Sanders2021}

    The longitudinal and transverse acoustic deformation potentials were found using equation \ref{eq:adp_2} with the polar angle between the phonon vector and the longitudinal axis of the $k_{z}$ conduction band minimum valley found at twenty evenly spaced angles from $0$ to $\pi$. For each polar angle, the azimuthal angle was varied and the average of the deformation potential found from eight evenly spaced points between 0 and $2\pi$.

    \begin{figure}[ht]
        \centering
        \includegraphics[width = 0.62\linewidth,angle=-90]{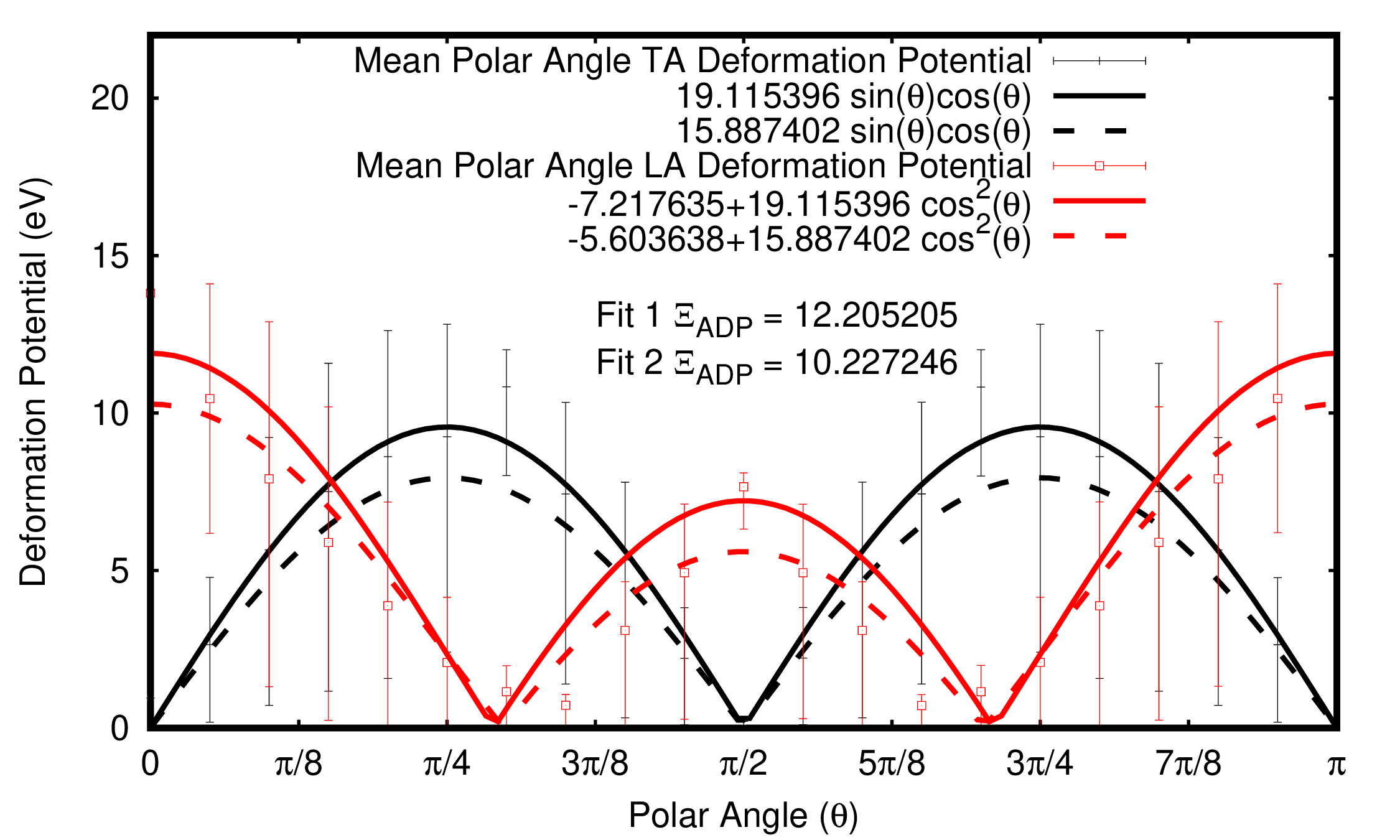}
        \caption[Polar Dependence of the Acoustic Deformation Potential in cBN.]{Polar dependence of the longitudinal (red) and transverse (black) acoustic deformation potential in the first conduction band valley of cBN calculated via quantum espresso (points) and equations \ref{eq:LA_angle} and \ref{eq:TA_angle}. The bars indicate the range of values found by varying the azimuthal angle for a given polar angle.}
        \label{fig:cBN_HV}
    \end{figure}
    Figure \ref{fig:cBN_HV} shows the angular dependence of the acoustic deformation potentials for the longitudinal, red, and the sum of the two transverse, black, acoustic phonon modes. The points show the average deformation potential for a given polar angle and the bars show the range of values calculated while varying the azimuthal angle for a given polar angle. As cBN is a polar material, the long range effects were subtracted before the deformation potential was calculated. The lines in figure \ref{fig:cBN_HV} show the fit of equations \ref{eq:LA_angle} and \ref{eq:TA_angle} to the calculated deformation potentials. The solid lines show the case where $\Xi_{u}$ and $\Xi_{d}$ were found separately by first fitting equation \ref{eq:TA_angle} to the average polar angle transverse acoustic deformation potential to find the value of $\Xi_{u}$ and then fixing this while fitting equation \ref{eq:LA_angle} to the average polar angle longitudinal acoustic deformation potential to find the value of $\Xi_{d}$. The uniaxial and dilatation deformation potentials were found to be $\Xi_{u}=19.12$eV and $\Xi_{d}=-7.22$eV.
    The dashed lines show the case where $\Xi_{u}$ and $\Xi_{d}$ were found together by fitting equation \ref{eq:LA_angle} to the average polar angle longitudinal acoustic deformation potential. The uniaxial and dilatation deformation potentials were found to be $\Xi_{u}=15.89$eV and $\Xi_{d}=-5.60$eV. The black dashed line is then the plot of equation \ref{eq:TA_angle} for $\Xi_{u}=16.4$eV.

    Unlike Si and diamond, cBN shows a large range of values for longitudinal and transverse deformation potentials due to the rotation about an azimuthal angle for a given polar angle. This could be due to the conduction band minimum being located at the $X$ symmetry point, as for a given polar angle the deformation potentials are calculated for 8 evenly spaced azimuthal angles, and so in half the cases the phonon vector has a component pointing either towards the $U$ high symmetry point or the $W$ high symmetry point. A clearer picture could then be achieved by calculating the deformation potential at more azimuthal angles or by picking azimuthal angles that don't point directly at other high symmetry points.

    Values for the longitudinal and transverse acoustic deformation potential can then be found using equations \ref{eq:LA_average} and \ref{eq:TA_average}. The contributions of the longitudinal and transverse acoustic modes can be combined into a single value using equation \ref{eq:ADP_contribution} with $v_{l} = 15.4\times 10^{3}$ m/s and $v_{t} = 11.8\times 10^{3}$ m/s \cite{Levinshtein2001}. The acoustic deformation potential found by fitting equation \ref{eq:TA_angle} to the mean polar angle transverse acoustic deformation potential to find $\Xi_{u}$ and then fitting \ref{eq:LA_angle} to the mean polar angle longitudinal acoustic deformation potential to find $\Xi_{d}$, referred to as fit 1, was $\Xi_{ADP}=12.21$eV. The acoustic deformation potential found by fitting equation \ref{eq:LA_angle} to the mean polar angle longitudinal acoustic deformation potential to find both $\Xi_{d}$ and $\Xi_{u}$, referred to as fit 2, was $\Xi_{ADP}=10.23$eV. The deformation potential calculated here does compare well with what is seen in the work of Siddiqua et al. \cite{Siddiqua2020}, however, they note that for the acoustic deformation potential, "reasonable estimates are employed" as the "particular material parameter has yet to be experimentally determined". It is unclear exactly how the value for the acoustic deformation potential was determined and so this similarity could be coincidental. 

    As with Si and diamond, the effective mass at the conduction band minimum was calculated by fitting a parabolic curve to the band structure parallel and perpendicular to the $\Delta$ direction in order to calculate the longitudinal and transverse effective masses. These were found to be $m_{l}^{*}=0.92m_{0}$ and $m_{t}=0.30m_{0}$ which gives a DOS effective mass of $m_{DOS}^{*}=0.44m_{0}$. The DOS effective mass compares well with the values of $m_{DOS}^{*}=0.413m_{0}$ \cite{Zhu2023} and $m_{DOS}^{*}=0.43m_{0}$ \cite{Siddiqua2020}, however, the latter presents values for the longitudinal and transverse effective masses that have a greater anisotropy, $m_{l}^{*}=1.2m_{0}$ and $m_{t}=0.26m_{0}$, calculated via first principles from reference \cite{Xu1991}. Most MC simulations use the DOS effective mass and assume the bands are isotropic and so in this case, this disagreement in the anisotropy wouldn't matter as there is good agreement in the DOS effective mass, however, if the MC simulation was intended to model valleytronics as in references \cite{Hammersberg2014,Djurberg2022} for diamond, the anisotropy of the effective masses would be important as electrons in perpendicular valleys would experience the acceleration of an electric field differently.

    The application of these \textit{ab initio} methods have allowed for the calculation of scattering parameters in a material that lacks experimental data from which empirical values can be calculated. The DOS effective mass is in good agreement with what has been employed previously in the literature \cite{Siddiqua2020,Zhu2023} however there is a slight disagreement in the anisotropy of the longitudinal and transverse effective mass as compared to values presented in reference \cite{Xu1991}. The deformation potential is also in good agreement with what is seen in reference \cite{Siddiqua2020} and even though this is a "reasonable estimate" the similarity in the DOS effective mass and the acoustic deformation potential means that the contribution of the acoustic phonon scattering will be similar using the two sets of parameters.

\section{\label{sec:conc}Conclusion}
  In this paper, DFT and DFPT have been employed to develop a method for calculating the acoustic deformation potential for MC simulation that improves upon the the work of Li et al. \cite{Li2021} and Yang et al. \cite{Yang2024}. This was achieved by fitting equations describing the transverse and longitudinal deformation potential to results calculated via first principles, and using these to determine the average acoustic deformation potential. The use of this method led to a prediction of the deformation potential in Si which was in excellent agreement with what has been seen previously, a deformation potential for diamond greater than what has been presented in the literature, and a deformation potential for cBN which supports the reasonable estimate employed by Siddiqua et al. \cite{Siddiqua2020}. The discrepancy in the deformation potential calculated for diamond, and the large range of values seen in the literature is thought to be due to their use as free parameters to fit the simulation output to experimental results. This is also impacted by the choice of the effective mass used in the acoustic phonon scattering rate and may also be impacted through the choice of the other parameters in equation \ref{eq:ADP_scatter}. As such, this range of values observed in diamond but not in Si could be because Si has been extensively studied as a semiconductor for decades, and so the other scattering parameters are better characterised giving less freedom to the deformation potential. Due to the agreement between the deformation potential calculated for Si and the values from literature, and the fact that deformation potential for diamond is just above the range of values reported; it is thought that the deformation potential calculated for cBN is a suitable value for use in MC simulation. Also, due to the fact that the DOS effective mass calculated here was in good agreement with what was used in references \cite{Siddiqua2020,Zhu2023}, the acoustic phonon scattering rate will have a similar effect on modelling of electrons when used in a MC simulation. The source of the large spread of values in the angular dependent deformation potential in cBN may be due to the anisotropic band structure in cBN at the $X$ symmetry point, where the conduction band minimum is located; however, it is not known what impact this azimuthal anisotropy has on the deformation potentials reported here for cBN.
  
  Future work will be conducted to employ these acoustic deformation potentials, along with other scattering parameters calculated from first principles, in a MC simulation of the Boltzmann transport equation for diamond and cBN. These will explore the use of these \textit{ab initio} scattering parameters in transport simulations to see how accurate the results are when compared to simulations using empirically calculated parameters and experimental results, in the case of diamond; or estimated scattering parameters in, the case of cBN.

\section{Acknowledgements}
  P. W. would like to acknowledge the support of UKRI via a DTP scholarship.

\bibliography{library}

\end{document}